\documentclass[aps,pra,twocolumn,superscriptaddress,nofootinbib]{revtex4-1}
\usepackage{amsmath}
\usepackage{amssymb}
\usepackage{graphicx}
\usepackage{mathrsfs}
\usepackage{ntheorem}
\usepackage{enumerate}
\usepackage{enumitem}
\usepackage{times,txfonts}
\usepackage{subfigure}
\newcommand{\ket}[1]{|#1\rangle}
\newcommand{\bra}[1]{\langle #1|}
\newcommand{\Tr}{\mathrm{Tr}}

\newcommand{\abs}[1]{\lvert #1\rvert}

\def\CC{{\rm\kern.24em \vrule width.04em height1.46ex depth-.07ex \kern-.30em C}}
\def\RR{{\rm\kern.24em \vrule width.04em height1.46ex depth-.07ex
\kern-.30em R}}
\def\P{{\rm I\kern-.25em P}}

\begin{document}
\title{Enhancing coherence of a state by stochastic strictly incoherent operations}
\author{C. L. Liu}
\affiliation{Department of Physics, Shandong University, Jinan 250100, China}
\author{Yan-Qing Guo}
\affiliation{Department of Physics, Dalian Maritime University, Dalian 116026, China}
\author{D. M. Tong}
\email{tdm@sdu.edu.cn}
\affiliation{Department of Physics, Shandong University, Jinan 250100, China}
\date{\today}
\begin{abstract}
In this paper, we address the issue of enhancing coherence of a state under stochastic strictly incoherent operations. Based on the $l_1$ norm of coherence, we obtain the maximal value of coherence that can be achieved  for a state undergoing  a stochastic strictly incoherent operation and the maximal probability of obtaining the maximal coherence. Our findings indicate that a pure state can be transformed into a maximally coherent state under a stochastic strictly incoherent operation  if and only if all the components of the pure state are nonzero while a mixed state can never be transformed into a maximally coherent state under a stochastic  strictly incoherent operation.
\end{abstract}
\maketitle

\section{Introduction}
Quantum coherence is a fundamental aspect of quantum physics, describing the capability of a quantum state to exhibit quantum interference phenomena. It is an essential component in quantum information processing \cite{Nielsen}, and plays a central role in emergent fields, such as quantum metrology \cite{Giovannetti,Giovannetti1}, nanoscale thermodynamics \cite{Aberg,Lostaglio,Lostaglio1}, and quantum biology \cite{Sarovar,Lloyd,Huelga,Lambert}.
Recently, quantification of coherence has attracted a growing interest due to the development of quantum information science \cite{Gour,Baumgratz,Marvian,Levi,Yuan,Aberg1,Streltsov,Bromley,Ma,Yao,Winter,Bagan,Napoli,Bera,Radhakrishnan,Ma,Streltsov1,Chitambar,Chitambar1,Chitambar2,
Yu,Yu1,Long,Fan,Guo,Marvian,Du,Zanardi,Streltsov2,Fan1,Vicente,Yadin,Girolami,Peng,Chiribella,Bu,Qi}.

By following the approach that has been established for entanglement resource \cite{Horodecki,Plenio}, Baumgratz et al. proposed a seminal framework for quantifying coherence \cite{Baumgratz}. It comprises four conditions, the coherence being zero (positive) for incoherent states (all other states), the monotonicity of coherence under incoherent operations, the monotonicity of coherence under selective measurements on average, and the nonincreasing of coherence under mixing of quantum states. The four conditions are fulfilled by a number of functionals of states, such as the $l_1$ norm of coherence and  the relative entropy of coherence, which can be taken as coherence measures. With these coherence measures, various topics of quantum coherence, such as the relations between quantum coherence and quantum correlations \cite{Ma,Streltsov}, the freezing phenomenon of coherence \cite{Bromley,Yu}, and the duality of coherence and path distinguishability \cite{Bagan,Bera}, have been investigated.

Quantum coherence is a useful physical resource in performing quantum information processing tasks. When a system is used to perform some task, it is often expected to have a sufficiently large quantity of coherence. In practical applications, we may need to enhance the coherence of a state. This may not be a difficult problem if we do not restrict the choices of operations, as there are many operations that can increase the coherence of a state. However, it will be a challenging topic if the operations are restricted to incoherent operations.

Investigations on this topic have been started in Ref. \cite{Yuan}, where a coherence distillation procedure for pure states under collective strictly incoherent operations was introduced. Recently, the coherence distillation of mixed states under collective incoherent operations was addressed in Ref. \cite{Winter}. It shows that a state with a smaller quantity of coherence can be asymptotically transformed into a state with a larger quantity of coherence under collective incoherent operations. However, the distillation procedure, as a scheme of enhancing coherence, requires the copies of states to be sufficiently large, and needs collective measurements on a large number of states, which are often very delicate as they involve controlled interaction among different particles. In the present paper, we consider an alternative scheme of enhancing coherence of a state under stochastic incoherent operations acting on a single state. Similar schemes have been used to enhancing entanglement of an individual pair of particles \cite{Linden,Kent,Kent1}.  We here focus our discussion on the widely used $l_1$ norm of coherence, and restrict the operations to  strictly incoherent operations, which are a physically well-motivated set of free operations for coherence and a strong candidate for free operations \cite{Yadin}. We will give the maximal value of coherence that can be achieved  for a state undergoing  a stochastic strictly incoherent operation and the maximal probability of obtaining the maximal coherence.

The paper is organized as follows. In Sec. II, we present some preliminaries. In Sec. III, we put forward the first theorem, which gives the maximal value of coherence that can be achieved by performing a stochastic strictly incoherent operation on a state. In Sec. IV, we put forward the second theorem, which gives the maximal probability of obtaining a coherence-enhanced state with the maximal coherence. In Sec. V, by applying the theorems to pure state and mixed states, respectively, we further put forward two corollaries.
Section VI is a summary of our findings.

\section{ Preliminaries}

Let $\mathcal {H}$ represent  the Hilbert space of a $d$-dimensional quantum system. A particular basis of $\mathcal {H}$ is denoted as $\{\ket{i}, ~i=1,2,\cdot\cdot\cdot,d\}$, which is chosen according to the physical problem under discussion. Coherence of a state is then measured based on the basis chosen \cite{Baumgratz}. Specifically, a state is said to be incoherent if it is diagonal in the basis.

The coherence effect of a state is ascribed to the off-diagonal elements of its density matrix with respect to the chosen basis. An intuitive measure of coherence is the $l_1$ norm of coherence. If we use $\rho=\sum_{i,j=1}^d\rho_{ij}\ket{i}\bra{j}$ to represent a general state, the $l_1$ norm of coherence is defined straightforwardly by the sum of absolute values of all the off-diagonal elements,
\begin{eqnarray}
C_{l_1}(\rho)=\sum_{i\neq j}|\rho_{ij}|.\label{L}
\end{eqnarray}
The $l_1$ norm of coherence is one of the most widely used measures in the resource theory of coherence.
It fulfills $0\leq C_{l_1}(\rho)\leq{d-1}$. The upper bound is attained only for the maximally coherent state, which has the form of $\ket{\psi^d_{\max}}=\frac1{\sqrt{d}}\sum_{i=1}^d e^{i\delta_i}\ket{i}$ with $\delta_i$ being real numbers.

To introduce the notion of stochastic strictly incoherent operations, we first recall strictly incoherent operations. A strictly incoherent operation is a completely positive trace-preserving map, expressed as $\Lambda(\rho)=\sum_n K_n\rho K_n^\dagger$,
where the Kraus operators $K_n$ satisfy not only  $\sum_n K_n^\dagger K_n= I$ but also $K_n\mathcal{I}K_n^\dagger\subset \mathcal{I}$ and $K_n^\dagger\mathcal{I}K_n\subset\mathcal{I}$ for $K_n$, i.e., each $K_n$ as well $K_n^\dagger$ maps an incoherent state to an incoherent state. Here, $\mathcal{I}$ represents the set of incoherent states.
There is at most one nonzero element in each column (row) of $K_n$, and such a $K_n$ is called a strictly incoherent Kraus operator.

With the aid of strictly incoherent operations, we may introduce the notion of stochastic strictly incoherent operations. A stochastic strictly incoherent operation is constructed by a subset of strictly incoherent Kraus operators. Without loss of generality, we denote the subset as $\{K_{1},K_{2},\dots, K_{L}\}$. Otherwise, we may renumber the subscripts of these Kraus operators. Then, a stochastic strictly incoherent operation, denoted as $\Lambda_s(\rho)$, is defined by
\begin{equation}
\Lambda_s(\rho)=\frac{\sum_{n=1}^L K_{n}\rho K_{n}^{\dagger}}{\Tr(\sum_{n=1}^LK_{n}\rho K_{n}^{\dagger})},
\label{lams}
\end{equation}
where $\{K_{1},K_{2},\dots, K_{L}\}$ satisfies $\sum_{n=1}^L K_{n}^{\dagger}K_{n}\leq I$. Similar notions on stochastic operations can be seen in previous works \cite{Bu,Aolita}. Clearly, the state $\Lambda_s(\rho)$ is obtained with probability $P=\Tr(\sum_{n=1}^LK_{n}\rho K_{n}^{\dagger})$ under a stochastic strictly incoherent operation $\Lambda_s$, while state $\Lambda(\rho)$ is fully deterministic under a strictly incoherent operation $\Lambda$.

It is known that a strictly incoherent operation does not increase coherence of a state, i.e., $C_{l_1}(\Lambda(\rho))\leq C_{l_1}(\rho)$, and there is always $\sum_n p_n C_{l_1}(\rho_n)\leq C_{l_1}(\rho)$, where $p_n=\Tr(K_n\rho K_n^\dagger)$,
$\rho_n=K_n\rho K_n^\dagger/\Tr(K_n\rho K_n^\dagger)$. However, these relations do not prevent us from obtaining probabilistically a state with larger coherence under a stochastic strictly incoherent operation. Namely, it is possible to have $C_{l_1}(\Lambda_s(\rho))>C_{l_1}(\rho)$ for a stochastic strictly incoherent operation $\Lambda_s$, although $C_{l_1}(\Lambda(\rho))\leq C_{l_1}(\rho)$ is always true for a strictly incoherent operation $\Lambda$. In fact, some of $\rho_n=K_n\rho K_n^\dagger/\Tr(K_n\rho K_n^\dagger)$, obtained under a strictly incoherent operation with selective measurements, may have a larger value of coherence than $\rho$. If we pick out only those $\rho_n$ satisfying $C(\rho_n)>C(\rho)$ and discard other $\rho_n$ with smaller $C(\rho_n)$, we may probabilistically obtain a mixed state ${\sum_n}_{C(\rho_n)>C(\rho)} p_n \rho_n$, which has a larger value of coherence. Then, a desired state is obtained with probability $P$ under a strictly incoherent operation with selective measurements. Therefore, we may enhance the coherence of a state by a stochastic strictly incoherent operation.

We are particularly interested in the maximal value of coherence that can be achieved when a state undergoes a stochastic strictly incoherent operation.

\section{Optimal coherence enhancement}

We take the $l_1$ norm of coherence as our measure of coherence. We aim to find the optimal coherence enhancement, i.e., the maximal value of coherence that can be obtained by performing a stochastic strictly incoherent operation on a state.

The state under consideration is denoted as $\rho=\sum_{ij}\rho_{ij}\ket{i}\bra{j}$. Based on it, we can define three matrices $|\rho|$, $\rho_d$, and $\rho_d^{-\frac12}$, where $|\rho|$ reads $|\rho|=\sum_{ij}|\rho_{ij}|\ket{i}\bra{j}$, $\rho_d=\sum_i\rho_{ii}\ket{i}\bra{i}$, and $\rho_d^{-\frac12}$ is a diagonal matrix with elements
\begin{equation}
 (\rho_d^{-\frac12})_{ii}=\left\{\begin{array}{ll} \rho_{ii}^{-\frac12}, &\text{if} ~ \rho_{ii}\neq0;\\
 0,&\text{if}~ \rho_{ii}= 0. \end{array}\right.
 \label{rho1/2}
 \end{equation}
Then, our main findings can be expressed as  the following theorems.

\emph{Theorem 1.} The maximal value of coherence that can be obtained by performing a stochastic strictly incoherent operation on $\rho$ reads
\begin{equation}
\max_{\Lambda_s}C_{l_1}\left(\Lambda_s(\rho)\right)=\lambda_{\max}(\rho_d^{-\frac12}\abs{\rho}\rho_d^{-\frac12})-1,
\label{theorem}
\end{equation}
where $\lambda_{\max}(\rho_d^{-\frac12}\abs{\rho}\rho_d^{-\frac12})$ represents the largest eigenvalue of the matrix $\rho_d^{-\frac12}\abs{\rho}\rho_d^{-\frac12}$.

We now prove the theorem.

First, we show that
\begin{equation}
C_{l_1}\left(\Lambda_s(\rho)\right)\leq \max_{K_n} C_{l_1}\left(\frac{K_n\rho K_n^{\dagger}}{\Tr(K_n\rho K_n^{\dagger})}\right),
\label{proof10}
\end{equation}
for any stochastic strictly incoherent operation defined as  Eq. (\ref{lams}), where $K_n\in\{K_{1},K_{2},\dots, K_{L}\}$.
To this end, we rewrite  Eq. (\ref{lams}) as
\begin{eqnarray}
\Lambda_s(\rho)=\frac{\sum_{n=1}^L K_n\rho K_n^{\dagger}}{\Tr(\sum_{n=1}^LK_n\rho K_n^{\dagger})}=\sum_{n=1}^L p_n\rho_n,
\label{proof11}
\end{eqnarray}
where $p_n=\Tr(K_n\rho K_n^{\dagger})/\Tr(\sum_{m=1}^L K_m\rho K_m^{\dagger})$ and $\rho_n= K_n\rho K_n^{\dagger}/\Tr(K_n\rho K_n^{\dagger})$.
Note that $K_n\rho K_n^{\dagger}$ is positive semidefinite and therefore $\Tr(K_n\rho K_n^{\dagger})\neq 0$ unless $K_n\rho K_n^{\dagger}=0$.
Since coherence is non-increasing under mixing of quantum states, we have
\begin{eqnarray}
C_{l_1}\left(\Lambda_s(\rho)\right)&&=C_{l_1}\left(\sum_{n=1}^L p_n\rho_n\right)\nonumber\leq \sum_{n=1}^L p_nC_{l_1}\left(\rho_n\right) \nonumber\\
&&\leq\max_{K_n} C_{l_1}\left(\frac{K_n\rho K_n^{\dagger}}{\Tr(K_n\rho K_n^{\dagger})}\right).
\label{proof12}
\end{eqnarray}
Equation (\ref{proof12}) immediately leads to Eq. (\ref{proof10}).

Second, we show that
\begin{equation}
\max_{K_n} C_{l_1}\left(\frac{K_n\rho K_n^{\dagger}}{\Tr(K_n\rho K_n^{\dagger})}\right)\leq \lambda_{\max}(\rho_d^{-\frac12}\abs{\rho}\rho_d^{-\frac12})-1.
\label{proof13}
\end{equation}
To this end, we only need to show that for any strictly incoherent Kraus operator $K$ with $K\rho K^{\dagger}\neq 0$, there is always
\begin{equation}
C_{l_1}\left(\frac{K\rho K^{\dagger}}{\Tr(K\rho K^{\dagger})}\right)\leq \lambda_{\max}(\rho_d^{-\frac12}\abs{\rho}\rho_d^{-\frac12})-1.
\label{proof14}
\end{equation}
Since there is at most one nonzero element in each column (row) of a strictly incoherent Kraus operator, any $K$ can always be transformed into a diagonal form via an incoherent unitary matrix, which does not change the value of $C_{l_1}\left(K\rho K^{\dagger}/\Tr(K\rho K^{\dagger})\right)$.
Hence, without loss of generality, we may let
\begin{equation}
K=\text{diag}(a_1,a_2,...,a_d), \label{Kraus}
\end{equation}
where $a_i$ are complex numbers.
We then have
\begin{eqnarray}
C_{l_1}\left(\frac{K\rho K^{\dagger}}{\Tr(K\rho K^{\dagger})}\right)&&=\frac{\sum_{i\neq j}\abs{a_i}\abs{a_j}\abs{\rho_{ij}}}{\sum_i\abs{a_i}^2\rho_{ii}}\nonumber\\&&=\frac{\sum_{i,j}\abs{a_i}\abs{a_j}\abs{\rho_{ij}}}{\sum_i\abs{a_i}^2\rho_{ii}}-1.
\label{relation1}
\end{eqnarray}
We further introduce a vector, i.e., a column matrix, $\ket{\varphi}=\frac{1}{\sqrt{\sum_i\abs{a_i}^2\rho_{ii}}}\rho_d^{\frac{1}{2}}(\abs{a_1}, \abs{a_2},...,\abs{a_n})^t$, which satisfies $\langle \varphi\ket{\varphi}=1$. Hereafter, we use $M^t$ to denote the transpose of matrix $M$. It is easy to verify that
\begin{equation}
\bra{\varphi}\rho_d^{-\frac12}\abs{\rho}\rho_d^{-\frac12}\ket{\varphi}=\frac{\sum_{i,j}\abs{a_i}\abs{a_j}\abs{\rho_{ij}}}{\sum_i\abs{a_i}^2\rho_{ii}},
\label{relation2}
\end{equation}
where $\abs{\rho}=\sum_{i,j}\abs{\rho_{ij}}\ket{i}\bra{j}$, and $\rho_d^{-\frac{1}{2}}$ is defined by Eq. (\ref{rho1/2}).  Indeed, by directly substituting the expressions of $\ket{\varphi}$,  $\abs{\rho}$ and  $\rho_d^{-\frac12}$ into  $\bra{\varphi}\rho_d^{-\frac12}\abs{\rho}\rho_d^{-\frac12}\ket{\varphi}$, Eq. (\ref{relation2}) can be obtained. Then, Eq. (\ref{relation1}) is written as
\begin{equation}
C_{l_1}\left(\frac{K\rho K^{\dagger}}{\Tr(K\rho K^{\dagger})}\right)=\bra{\varphi}\rho_d^{-\frac12}\abs{\rho}\rho_d^{-\frac12}\ket{\varphi}-1.
\label{relation3}
\end{equation}
Note that $\bra{\varphi}\rho_d^{-\frac12}\abs{\rho}\rho_d^{-\frac12}\ket{\varphi}$ can be regarded as the average value of the matrix operator $\rho_d^{-\frac12}\abs{\rho}\rho_d^{-\frac12}$ with respect to the vector $\ket{\varphi}$. There is   $\bra{\varphi}\rho_d^{-\frac12}\abs{\rho}\rho_d^{-\frac12}\ket{\varphi} \leq \lambda_{\max}(\rho_d^{-\frac12}\abs{\rho}\rho_d^{-\frac12})$, where  $\lambda_{\max}(\rho_d^{-\frac12}\abs{\rho}\rho_d^{-\frac12})$ represents the largest eigenvalue of $\rho_d^{-\frac12}\abs{\rho}\rho_d^{-\frac12}$.
Thus, we obtain the expression,  $C_{l_1}\left(K\rho K^{\dagger}/\Tr(K\rho K^{\dagger})\right)\leq \lambda_{\max}(\rho_d^{-\frac12}\abs{\rho}\rho_d^{-\frac12})-1 $, i.e., Eq. (\ref{proof14}), which naturally implies Eq. (\ref{proof13}).

Third, we show that for any state $\rho$, there always exists a strictly incoherent Kraus operator $K^\prime$, which satisfies
 \begin{equation}
C_{l_1}\left(\frac{K^\prime\rho K^{\prime^{\dagger}}}{\Tr(K^\prime\rho K^{\prime^{\dagger}})}\right)= \lambda_{\max}(\rho_d^{-\frac12}\abs{\rho}\rho_d^{-\frac12})-1.
\label{proof15}
\end{equation}
To this end, we use  $\ket{\varphi_{\max}}=\left(\varphi_1, \varphi_2,\dots, \varphi_d\right)^t$ to denote the normalized eigenvector corresponding to the largest eigenvalue of $\rho_d^{-\frac12}\abs{\rho}\rho_d^{-\frac12}$. That is,  $\rho_d^{-\frac12}\abs{\rho}\rho_d^{-\frac12}\ket{\varphi_{\max}}= \lambda_{\max}(\rho_d^{-\frac12}\abs{\rho}\rho_d^{-\frac12})  \ket{\varphi_{\max}}$.
Noting that every component of the eigenvector corresponding to the largest eigenvalue of a nonnegative matrix can be chosen to be nonnegative \cite{Horn,Cooper}, we can take all $\varphi_i$ to be nonnegative numbers. With the help of  $\ket{\varphi_{\max}}$, it is easy to find a strictly incoherent Kraus operator satisfying Eq. (\ref{proof15}). For instance, such a strictly incoherent Kraus operator can be taken as
\begin{equation}
K^\prime=k~U_{in}~\text{diag}(a^{\prime}_1,a^{\prime}_2,...,a^{\prime}_d),
\label{Krausprime}
\end{equation}
where
\begin{equation}
 a^{\prime}_i=\left\{\begin{array}{ll}\frac{\varphi_i}{\sqrt{\rho_{ii}}},& \text{if} ~ \rho_{ii}\neq 0,\\
 0,&\text{if}~ \rho_{ii}= 0, \end{array}\right.\nonumber
\end{equation}
$U_{in}$ is an arbitrary incoherent unitary matrix, and  $k$ is a complex number for guaranteeing ${K^\prime}^\dagger K^\prime\leq I$.
In this case, we have
\begin{equation}
\begin{split}
C_{l_1}\left(\frac{K^\prime\rho K^{\prime^{\dagger}}}{\Tr(K^\prime\rho K^{\prime^{\dagger}})}\right)={\sum_{i,j}}^{\text{ex}}\varphi_i\rho_{ii}^{-\frac{1}{2}} \abs{\rho_{ij}}\rho_{jj}^{-\frac{1}{2}}\varphi_j-1.
\end{split}
\label{proof17}
\end{equation}
Here, the superscript ``ex'' in $\sum_{i,j}^{\text{ex}}$ means that the sum excludes the terms with $\rho_{ii}=0$. Since $\sum_{i,j}^{ex}\varphi_i\rho_{ii}^{-\frac{1}{2}} \abs{\rho_{ij}}\rho_{jj}^{-\frac{1}{2}}\varphi_j$ can be written as $\bra{\varphi_{\max}}\rho_d^{-\frac12}\abs{\rho}\rho_d^{-\frac12}\ket{\varphi_{\max}}$, we have, from Eq. (\ref{proof17}),
\begin{eqnarray}
C_{l_1}\left(\frac{K^\prime\rho K^{\prime^{\dagger}}}{\Tr(K^\prime\rho K^{\prime^{\dagger}})}\right)&&=\bra{\varphi_{\max}}\rho_d^{-\frac12}\abs{\rho}\rho_d^{-\frac12}\ket{\varphi_{\max}}-1\nonumber\\&&=  \lambda_{\max}(\rho_d^{-\frac12}\abs{\rho}\rho_d^{-\frac12})-1,
\label{proof18}
\end{eqnarray}
i.e., Eq. (\ref{proof15}).

From Eqs. (\ref{proof10}), (\ref{proof13}) and (\ref{proof15}), we immediately obtain Eq. (\ref{theorem}). This completes the proof of \emph{Theorem 1}.

\section{Maximal probability of optimal coherence enhancement}

In this section, we investigate the probability of optimal coherence enhancement.  For a given state $\rho$ undergoing a stochastic strictly incoherent operation $\Lambda_s$, the probability of obtaining the state $\Lambda_s(\rho)$ reads
$P=\Tr(\sum_{n=1}^LK_n\rho K_n^{\dagger})$.
In the case of optimal coherence enhancement, $\Lambda_s$ is defined only by one Kraus operator $K^\prime$, and the probability is reduced to
\begin{equation}
P=\Tr(K^\prime\rho K^{\prime\dagger}).
\label{prob}
\end{equation}
We aim to calculate the probability of obtaining the maximal enhanced state $\Lambda_s(\rho)$.

First, we consider the case of $\rho$ being irreducible. That is, $\rho$ cannot be transformed into a block diagonal matrix only by using a permutation matrix.  Since $\rho$ is irreducible, the matrix $\rho_d^{-\frac12}\abs{\rho}\rho_d^{-\frac12}$ is irreducible, too. Then, according to the Perron-Frobenius's theorem \cite{Horn}, there exists a unique eigenvector of $\rho_d^{-\frac12}\abs{\rho}\rho_d^{-\frac12}$ corresponding to the maximal eigenvalue $\lambda_{\max}(\rho_d^{-\frac12}\abs{\rho}\rho_d^{-\frac12})$ such that all the components of the eigenvector are positive.
We still use  $\ket{\varphi_{\max}}=\left(\varphi_1, \varphi_2,\dots, \varphi_d\right)^t$ to denote the normalized eigenvector corresponding to the largest eigenvalue, where $\varphi_i>0$ for all $i=1,...,d$.
For the irreducible density matrix $\rho$, the general form of the optimal strictly incoherent Kraus operator can be written as
\begin{equation}
K^\prime=k~U_{in}~\text{diag}\left(\frac{\varphi_1}{\sqrt{\rho_{11}}},\frac{\varphi_2}{\sqrt{\rho_{22}}},
\cdots,\frac{\varphi_d}{\sqrt{\rho_{dd}}}\right),\label{irrk}
\end{equation}
where $U_{in}$ is an arbitrary incoherent unitary matrix, and $k$ satisfies
\begin{equation}
|k|\leq \min_i\frac{\sqrt{\rho_{ii}}}{\varphi_i},\label{absk}
\end{equation}
due to the requirement $K^{\prime\dagger} K^\prime\leq I$.

Substituting Eq. (\ref{irrk}) into  $P=\Tr(K^\prime\rho K^{\prime\dagger})$, we immediately have
\begin{equation}
P=|k|^2.\label{pk2}
\end{equation}

From Eqs. (\ref{absk}) and (\ref{pk2}), we can obtain the maximal probability,
\begin{equation}
P_{\max}=\min_i\frac{\rho_{ii}}{\varphi_i^2}.
\label{pk2b}
\end{equation}
The corresponding optimal Kraus operator is given by Eq. (\ref{irrk}) with $|k|=\min_i\frac{\sqrt{\rho_{ii}}}{\varphi_i}$.

Second, we consider the case of $\rho$ being reducible. $\rho$ is said to be reducible if it can be transformed into a block diagonal matrix only by using a permutation matrix $M$. Since any permutation matrix is an incoherent unitary and the coherence of a state is invariant under an incoherent unitary, the states $\rho$ and $M\rho M^t$ have the same coherence. Furthermore, there is  $\max_{\Lambda_s}C_{l_1}\left(\Lambda_s(\rho)\right)=\max_{\Lambda_s}C_{l_1}\left(\Lambda_s(M\rho M^t)\right)$, which implies that $P_{\max}(\rho)=P_{\max}(M\rho M^t)$. Therefore, we only need to consider the case of $\rho=p_1\rho_1 \oplus p_2\rho_2 \oplus\cdots\oplus p_n\rho_n \oplus \textbf{0 }$, where each $\rho_\alpha=\sum_{i,j}\rho_{ij}^\alpha\ket{i}\bra{j} $ ($\alpha=1,2,\cdots,n$) is an irreducible density operator defined on the $d_\alpha$-dimensional subspace $\mathcal {H}_\alpha$, $p_\alpha>0$ satisfies $\sum_{\alpha=1}^np_\alpha=1$, and $\textbf{0}$ represents a square matrix of dimension $d_0=d-(d_1+d_2+\cdots +d_n)$ with all its elements being zero.
In this case, we have $\abs{\rho}=p_1\abs{\rho_1} \oplus p_2\abs{\rho_2}  \oplus\cdots\oplus p_n\abs{\rho_n} \oplus \textbf{0 }$ and $\rho_d^{-\frac12}=(p_1\rho_d)_1^{-\frac12}  \oplus (p_2\rho_d)_2^{-\frac12}  \oplus\cdots\oplus (p_n\rho_d)_n^{-\frac12}  \oplus \textbf{0 }$. Then, $\rho_d^{-\frac12}\abs{\rho}\rho_d^{-\frac12}$ can be expressed as
\begin{equation}
\rho_d^{-\frac12}\abs{\rho}\rho_d^{-\frac12}=A_1 \oplus A_2  \oplus\cdots\oplus A_n \oplus \textbf{0 },
\label{theorem21}
\end{equation}
where each $A_\alpha=(\rho_d)_\alpha^{-\frac12}\abs{\rho_\alpha}(\rho_d)_\alpha^{-\frac12} $  is an irreducible nonnegative matrix.

We use  $\lambda_{\max}^\alpha$ to denote the maximal eigenvalue of $A_\alpha$ and  $\ket{\varphi_{\max}^\alpha}=\left(\varphi^\alpha_1, \varphi^\alpha_2,\dots, \varphi^\alpha_{d_\alpha}\right)^t$ to denote the normalized eigenvector of $A_\alpha$ corresponding to the eigenvalue $\lambda_{\max}^\alpha$. Without loss of generality, we assume that $\lambda_{\max}^1\geq \lambda_{\max}^2 \geq \cdots\geq \lambda_{\max}^n$. Otherwise, we may rearrange the matrices $A_1, A_2, \cdots, A_n$ by a permutation transformation such that $\lambda_{\max}^i\geq \lambda_{\max}^{i+1} $. Since each $A_\alpha$ is an irreducible nonnegative matrix,   $\ket{\varphi_{\max}^\alpha}$ is unique if all its components are positive. Clearly,  $\lambda_{\max}^1, \lambda_{\max}^2, \cdots, \lambda_{\max}^n$ are also eigenvalues of $\rho_d^{-\frac12}\abs{\rho}\rho_d^{-\frac12}$, and the maximal eigenvalue of $\rho_d^{-\frac12}\abs{\rho}\rho_d^{-\frac12}$ is given by $\lambda_{\max}=\max\{\lambda_{\max}^1, \lambda_{\max}^2, \cdots, \lambda_{\max}^n\}$.  Further, we suppose the degenerate degree of $\lambda_{\max}$ is $n_d$, i.e., $\lambda_{\max}^1= \lambda_{\max}^2=\cdots= \lambda_{\max}^{n_d}=\lambda_{\max}$. Then, the normalized eigenvectors of  $\rho_d^{-\frac12}\abs{\rho}\rho_d^{-\frac12}$ can be generally written as
\begin{equation}
\ket{\varphi_{\max}}=c_1 \ket{\varphi_{\max}^1}\oplus c_2\ket{\varphi_{\max}^2}\oplus\cdots\oplus c_{n_d}\ket{\varphi_{\max}^{n_d}}\oplus \ket{\textbf{0}},
\label{psimax}
\end{equation}
where $\ket{\textbf{0}}=(0,0,\cdots,0)^t$ is zero vector  and $c_i$ are the coefficients satisfying  $\sum_i^{n_d}  |c_i|^2=1$.
Here, $c_\alpha \ket{\varphi_{\max}^\alpha}\oplus c_\beta\ket{\varphi_{\max}^\beta}$  is defined as $\left(c_\alpha\varphi^\alpha_1, c_\alpha\varphi^\alpha_2,\dots, c_\alpha\varphi^\alpha_{d_\alpha}, c_\beta\varphi^\beta_1, c_\beta\varphi^\beta_2,\dots, c_\beta\varphi^\beta_{d_\beta} \right)^t$.

With these knowledge, it is easy to understand that the optimal strictly incoherent Kraus operator for reducible $\rho$ can be generally written as
\begin{equation}
K^\prime=U_{in}~(K^\prime_1\oplus K^\prime_2\oplus \cdot\cdot\cdot \oplus K^\prime_d\oplus\textbf{0}),
\label{rk}
\end{equation}
where $U_{in}$ is an arbitrary unitary incoherent operator, and
\begin{equation}
K^\prime_\alpha=k_\alpha~\text{diag}\left(\frac{\varphi^{\alpha}_1}{\sqrt{\rho^\alpha_{11}}},\frac{\varphi^{\alpha}_2}{\sqrt{\rho^\alpha_{22}}}, \cdot\cdot\cdot, \frac{\varphi^{\alpha}_{d_\alpha}}{\sqrt{\rho^\alpha_{d_\alpha,d_\alpha}}}\right)
\label{rk2}
\end{equation}
with $k_\alpha$ satisfying
\begin{equation}
|k_\alpha|\leq \min_i\frac{\sqrt{\rho^\alpha_{ii}}}{\varphi^{\alpha}_i}\label{pk2c}
\end{equation}
due to the requirement of $K^{\prime\dagger}_\alpha K^\prime_\alpha\leq I$.

Substituting Eqs. (\ref{rk}) and (\ref{rk2}) into $P=\Tr(K^\prime\rho K^{\prime\dagger})$, we have
\begin{equation}
P=\sum_\alpha p_\alpha\Tr K^\prime_\alpha \rho_\alpha K^\prime_\alpha=\sum_\alpha p_\alpha |k_\alpha|^2.
\label{rkP}
\end{equation}
From Eqs. (\ref{pk2c}) and (\ref{rkP}),  we obtain the maximal probability,
\begin{equation}
P_{\max}=\sum_\alpha p_\alpha \min_i\frac{\rho^\alpha_{ii}}{(\varphi^{\alpha}_i)^2}.
\label{rkP2}
\end{equation}
The corresponding optimal Kraus operator is given by Eqs. (\ref{rk}) and (\ref{rk2}) with $|k_\alpha|=\min_i\frac{\sqrt{\rho^\alpha_{ii}}}{\varphi^{\alpha}_i}$.

Clearly, the result for $\rho$ being irreducible can be taken as a special case of that for $\rho$ being reducible. Let $P_{\max}(\rho)$ represent the maximal probability at which the coherence of state $\rho$ can be enhanced to the maximal value by using stochastic strictly incoherent operations. We then can summarize the above results as \emph{Theorem 2}.

\emph{Theorem 2}. If $\rho$ is irreducible, then $P_{\max}(\rho)=\min_i\frac{\rho_{ii}}{\varphi_i^2}$, where $\varphi_{i}$ is the $i-$th component of the positive eigenvector $\ket{\varphi_{\max}}$ corresponding to the maximal eigenvalue of $\rho_d^{-\frac12}\abs{\rho}\rho_d^{-\frac12}$. If $\rho$ is reducible, i.e., it can be transformed by a permutation matrix into $p_1\rho_1 \oplus p_2\rho_2  \oplus\cdots\oplus p_n\rho_n \oplus \textbf{0 }$ with $\rho_\alpha$ being irreducible, then $P_{\max}={\sum_\alpha}_{(\lambda^\alpha_{\max}=\lambda_{\max})} p_\alpha P_{\max}(\rho_\alpha)$.

Note that the sum is only for those indexes $\alpha$ satisfying $\lambda_{\max}^\alpha=\lambda_{\max}$. The optimal strictly incoherent Kraus operator achieving the maximal probability can be generally expressed as Eq. (\ref{rk}), i.e., $K^\prime=U_{in}~(K^\prime_1\oplus K^\prime_2\oplus \cdot\cdot\cdot \oplus K^\prime_d\oplus\textbf{0})$, with
\begin{equation}
K_\alpha^\prime=\min_i\left(\frac{\sqrt{\rho^\alpha_{ii}}}{\varphi^{\alpha}_1}\right)\text{diag}\left(\frac{\varphi^{\alpha}_1}{\sqrt{\rho^\alpha_{11}}},\frac{\varphi^{\alpha}_2}{\sqrt{\rho^\alpha_{22}}}, \cdot\cdot\cdot, \frac{\varphi^{\alpha}_{d_\alpha}}{\sqrt{\rho^\alpha_{d_\alpha,d_\alpha}}}\right).
\label{rk3}
\end{equation}

Before going further, we give a simple example to illustrate the above theorems. Let us consider a system of single qubit, In the basis $\{\ket{0},\ket{1}\}$, a state of the qubit system can be generally expressed as
 $\rho=\frac12\left(
  \begin{array}{ccc}
    1+r\cos\theta & e^{-i\varphi}r\sin\theta\\
   e^{i\varphi}r\sin\theta& 1-r\cos\theta\\
  \end{array}
\right)$, where the parameters satisfy $0< r\leq 1$, $0< \theta < \pi$, and $0\leq \varphi \leq 2\pi$ for coherent states. For this state, we have
$\rho^{-\frac12}_d\abs{\rho} \rho^{-\frac12}_d=\left(
\begin{array}{ccc}
1&\frac{r\abs{\sin\theta}}{\sqrt{1-r^2\cos^2\theta}}\\
\frac{r\abs{\sin\theta}}{\sqrt{1-r^2\cos^2\theta}}& 1\\
\end{array}
\right)$, of which the largest eigenvalue and the corresponding eigenvector are $\lambda_{\max}(\rho_d^{-\frac12}\abs{\rho}\rho_d^{-\frac12})=1+\frac{r\abs{\sin\theta}}{\sqrt{1-r^2\cos^2\theta}}$ and
$\ket{\varphi_{\max}}= \frac{1}{\sqrt{2}}(1,1)^t$, respectively.  By \emph{Theorem 1}, we immediately obtain the optimal coherence enhancement,    $\max_{\Lambda_s}C_{l_1}\left(\Lambda_s(\rho)\right)= \frac{r\abs{\sin\theta}}{\sqrt{1-r^2\cos^2\theta}}$, which is obviously greater than or equal to   $C_{l_1}(\rho)=r\abs{\sin\theta}$. By \emph{Theorem 2}, we can obtain
the maximal probability of obtaining the optimal coherence enhancement,
$P_{\max}(\rho)=1-r\abs{\cos\theta}$.
The optimal strictly incoherent Kraus operator achieving the maximal probability can be generally expressed as  $K^{'}=\sqrt{1-r\abs{\cos\theta}}U_{in}~\text{diag}(\frac{1}{\sqrt{1+r\cos\theta}},\frac{1}{\sqrt{1-r\cos\theta}})$
with $U_{in}$ being an arbitrary $2\times2$ incoherent unitary matrix.

\section{Discussions}

In the previous sections, we have proved two theorems, of which one gives the maximal coherence that can be achieved by performing a stochastic strictly incoherent operation $\Lambda_s$ on a state $\rho$ and the other gives the maximal probability of obtaining the state $\Lambda_s(\rho)$ with the maximal coherence. We now make some further discussions by applying the theorems to pure states and mixed states, respectively. From them, we can infer the following two corollaries.

\emph{Corollary 3}.  A pure state $\ket{\phi}=(\phi_1,\phi_2,\dots, \phi_d)^t$ can be transformed into a maximally coherent state by a stochastic strictly incoherent operations  if and only if all the components $\phi_i$ are nonzero. The maximal probability of obtaining the maximally coherent state is $P_{\max}(\rho)=d\cdot\min_i\abs{\phi_i}^2$.

To derive \emph{Corollary 3} from \emph{Theorem 1} and \emph{Theorem 2}, we first calculate the largest eigenvalue of matrix $\rho_d^{-\frac12}\abs{\rho}\rho_d^{-\frac12}$ with $\rho=\ket{\phi}\bra{\phi}$ and $\ket{\phi}=(\phi_1,\phi_2,\dots, \phi_d)^t$. For an arbitrary pure state $\ket{\phi}$,  there are $\rho_{ij}=\phi_i\phi_j^*$. We then have
\begin{equation}
(\rho_d^{-\frac12}\abs{\rho}\rho_d^{-\frac12})_{ij}=\left\{\begin{array}{ll} 1,& \text{if} ~ \phi_i\phi_j^*\neq 0,\\
 0,&\text{if}~ \phi_i\phi_j^*=0.\nonumber
 \label{rhodij}
 \end{array}\right.
\end{equation}
That is, all the elements of matrix $\rho_d^{-\frac12}\abs{\rho}\rho_d^{-\frac12}$ are $1$ except for some rows and columns with zero elements. The maximal eigenvalue of such a matrix is equal to the number of the rows with elements $1$, i.e., the number of nonzero $\phi_i$, denoted as $r$.  Therefore, we have $\lambda_{\max}(\rho_d^{-\frac12}\abs{\rho}\rho_d^{-\frac12})=r$. From \emph{Theorem 1}, we immediately have  $\max_{\Lambda_s}C_{l_1}\left(\Lambda_s(\rho)\right)=r-1$.

If all the components of $\ket{\phi}$ are nonzero, there will be $r=d$ and therefore $\max_{\Lambda_s}C_{l_1}\left(\Lambda_s(\rho)\right)=d-1$, which means that  $\Lambda_s(\rho)$ is a maximally coherent state. In this case, each element of $\rho_d^{-\frac12}\abs{\rho}\rho_d^{-\frac12}$ is equal to  $1$, and the eigenvector corresponding to the largest eigenvalue of $\rho_d^{-\frac12}\abs{\rho}\rho_d^{-\frac12}$  is  $\ket{\varphi_{\max}}=\frac{1}{\sqrt{d}}\left(1, 1,\dots, 1\right)^t$. Hence, by using  \emph{Theorem 2}, we have  $P_{\max}(\ket{\phi}\bra{\phi})=d\cdot\min_i\abs{\phi_i}^2$.

\emph{Corollary 4.} A mixed state $\rho$ can never be transformed into a maximally coherent state by a stochastic  strictly incoherent operation.

To prove \emph{Corollary 4}, we only need to demonstrate that $\lambda_{\max}(\rho_d^{-\frac12}\abs{\rho}\rho_d^{-\frac12})<d$ for any mixed state $\rho=\sum_{ij}\rho_{ij}\ket{i}\bra{j}$.  By substituting the definitions of $\abs{\rho}$ and $\rho_d^{-\frac12}$ into $\rho_d^{-\frac12}\abs{\rho}\rho_d^{-\frac12}$, we can obtain
\begin{equation}
(\rho_d^{-\frac12}\abs{\rho}\rho_d^{-\frac12})_{ij}=\left\{\begin{array}{ll}\frac{|\rho_{ij}|}{\sqrt{\rho_{ii}\rho_{jj}}},& \text{if} ~ \rho_{ii}\rho_{jj}\neq 0;\\
 0,&\text{if}~ \rho_{ii}\rho_{jj}= 0.
 \label{rhodij}
 \end{array}\right.
\end{equation}
According to the Ger\v{s}gorin disk theorem \cite{Horn}, which implies that the largest eigenvalue of a square matrix $A$ with elements $A_{ij}$ is not larger than $\max_{i}\sum_{j}\abs{A_{ij}}$, we have
\begin{eqnarray}
\lambda_{\max}(\rho_d^{-\frac12}\abs{\rho}\rho_d^{-\frac12})&&\leq\max_i\sum_j \abs{\rho_d^{-\frac12}\abs{\rho}\rho_d^{-\frac12}}_{ij}
\nonumber\\
&&={\max_i}_{(\rho_{ii}\neq 0)}{\sum_j}^{ex}
\frac{|\rho_{ij}|}{\sqrt{\rho_{ii}\rho_{jj}}}.
\end{eqnarray}
Here, the superscript ``ex'' in $\sum_{j}^{\text{ex}}$ means that the sum excludes the terms with $\rho_{jj}=0$.
Since $\rho$ is a positive semidefinite matrix,  there is $\abs{\rho_{ij}}\leq \sqrt{\rho_{ii}\rho_{jj}}$. We then have  $\sum_j^{ex}
\frac{|\rho_{ij}|}{\sqrt{\rho_{ii}\rho_{jj}}}\leq d$ and therefore
\begin{eqnarray}
\lambda_{\max}(\rho_d^{-\frac12}\abs{\rho}\rho_d^{-\frac12})\leq d.
\end{eqnarray}

We now demonstrate that $\lambda_{\max}(\rho_d^{-\frac12}\abs{\rho}\rho_d^{-\frac12})$ cannot be equal to $d$ for a mixed state $\rho$. Otherwise, $\rho$ must be a pure state. From Eq. (\ref{rhodij}), we see that $(\rho_d^{-\frac12}\abs{\rho}\rho_d^{-\frac12})_{ij}\leq 1$. If $\lambda_{\max}(\rho_d^{-\frac12}\abs{\rho}\rho_d^{-\frac12})$ is assumed to be $d$, there must exists at least one row of matrix $\rho_d^{-\frac12}\abs{\rho}\rho_d^{-\frac12}$, in which all the elements are equal to $1$. However, the requirement that all the elements in one row of $\rho_d^{-\frac12}\abs{\rho}\rho_d^{-\frac12}$ are equal to $1$ will necessarily result in $(\rho_d^{-\frac12}\abs{\rho}\rho_d^{-\frac12})_{ij}=1$ for all $i$ and $j$ (See the Appendix for details). In this case, $\abs{\rho}$ must be a pure state, which further leads to the fact that $\rho$ is a pure state, too. Hence, $\lambda_{\max}(\rho_d^{-\frac12}\abs{\rho}\rho_d^{-\frac12})$ cannot be equal to $d$ for a mixed state. This completes the proof of \emph{Corollary 4}.
\section{Summary}

Quantum coherence is a useful physical resource, describing the abilities of a quantum system to perform quantum information processing tasks. While any incoherent operation cannot increase the coherence of a state, it does not prevent us from enhancing the coherence of a state by a stochastic incoherent operation. This paper addressed the topic of enhancing the coherence of a state by using a stochastic coherent operation. Considering that strictly incoherent operations are a physically well-motivated set of incoherent operations and therefore a strong candidate for incoherent operations, we have restricted our operations to the strictly incoherent operations. Based on the $l_1$ norm of coherence, we have investigated the possibility of enhancing the $l_1$ norm of coherence of a state by using a stochastic strictly incoherent operation.

Our main findings are presented as two theorems. \emph{Theorem 1} gives the maximal coherence that can be achieved by performing a stochastic strictly incoherent operation $\Lambda_s$ on a state $\rho$, while \emph{Theorem 2} gives the maximal probability of obtaining the state $\Lambda_s(\rho)$ with the maximal coherence.  It is shown that the maximal value of coherence is determined by the largest eigenvalue of the matrix  $\rho_d^{-\frac12}\abs{\rho}\rho_d^{-\frac12}$ while the maximal probability is determined by the eigenvectors corresponding to the largest eigenvalue.

As an application of the theorems, we further specify our discussions on pure states and mixed states, respectively, and we find that
a pure state can be transformed into a maximally coherent state by a stochastic strictly incoherent operations  if and only if all the components of the pure state are nonzero while a mixed state can never be transformed into a maximally coherent state by a stochastic  strictly incoherent operation.

In passing, we would like to point out that strictly incoherent Kraus operators can always be constructed by the system interacting with an ancilla and a general experimental setting has been suggested based on an interferometer in Ref. \cite{Yadin}. Thus, our scheme of enhancing coherence of a state may be experimentally demonstrated.

\begin{acknowledgments}
C.L.L. acknowledges support from the National Natural Science Foundation of China through Grant No. 11775129. Y.Q.G. acknowledges support from the Fundamental Research Funds for the Central Universities of China through Grant No. 3132015149. D.M.T. acknowledges support from the National Natural Science Foundation of China through Grant No. 11575101 and the National Basic Research Program of China through Grant No. 2015CB921004.
\end{acknowledgments}

\section*{APPENDIX}

We now show that the requirement that all the elements in one row of $\rho_d^{-\frac12}\abs{\rho}\rho_d^{-\frac12}$ are equal to $1$ will necessarily result in $(\rho_d^{-\frac12}\abs{\rho}\rho_d^{-\frac12})_{ij}=1$ for all $i$ and $j$.
For simplicity, we use $A_\text{d}$ to denote $\rho_d^{-\frac12}\abs{\rho}\rho_d^{-\frac12}$. $A_\text{d}$ is a $d$-dimensional positive semidefinite matrix, of which the elements $a_{ij}= (\rho_d^{-\frac12}\abs{\rho}\rho_d^{-\frac12})_{ij}$ satisfy $0\leq a_{ij}\leq1$ and $a_{ij}=a_{ji}\leq \sqrt{a_{ii}a_{jj}}$. We aim to show the proposition that all the elements $a_{ij}$, $i,j=1,2,\cdots,d$, are equal to $1$ if  $a_{i1}=a_{i2}=\cdots=a_{id}=1$ for some $i$. Without lost of generality, we let $i=1$, i.e.,  $a_{11}=a_{12}=\cdots=a_{1d}=1$.

It is obvious that the proposition is true for $d=2$. We assume that the proposition is valid for $d=n-1$, i.e.,  for the $(n-1)-$dimensional matrix $A_\text{n-1}$, all the elements $a_{ij}$, $i,j=1,2,\cdots,n-1$, are equal to $1$ if  $a_{11}=a_{12}=\cdots=a_{1,n-1}=1$.  Then we prove that it is also valid for $d=n$.

First, since $a_{1j}\leq \sqrt{a_{11}a_{jj}}$ and $a_{11}=a_{1j}=1$, there must be $a_{jj}=1$ for all $j=1,2,\cdots, n$. $A_n$ can be written as
\begin{equation}
A_\text{n}= \left(
\begin{array}{ccc}
A_\text{n-1} & C\\
C^\dag & 1\\
\end{array}
\right),\label{matrix_A}
\end{equation}
where $(A_\text{n-1})_{ij}=(A_\text{n})_{ij}=a_{ij}$, $i,j=1,2,\cdots, n-1$, and $C$  is a column matrix defined as $C=(1,a_{2,n},a_{3,n}\cdots,a_{n-1,n})^t$. Specially, $(A_\text{n-1})_{1i}=1$ for $i=1,2,\cdots,n-1$.

Second, we let $T= \left(\begin{array}{ccc} I_\text{n-1} & -C\\ \textbf{0}& 1\\  \end{array} \right)$.  As $A_\text{n}$ is a positive semidefinite matrix, the matrix $TA_\text{n}T^\dag$ must be a positive semidefinite matrix, too. Its explicit expression reads
\begin{equation}
  TA_\text{n}T^\dag= \left(
  \begin{array}{ccc}
    A_\text{n-1}- C C^\dag & \textbf{0}\\
    \textbf{0}& 1\\
  \end{array}
\right).\label{TMT}
\end{equation}
Equation (\ref{TMT}) implies that $TA_\text{n}T^\dag$ as well as $A_\text{n}$ is positive semidefinite if and only if $A_\text{n-1}- C C^\dag$ is positive semidefinite. The fact that $A_\text{n}$ is a positive semidefinite matrix necessarily leads to $A_\text{n-1}- C C^\dag$  being positive semidefinite, too.

Third, since $(A_\text{n-1})_{11}=(A_\text{n-1})_{12}=\cdots=(A_\text{n-1})_{1,n-1}=1$, all the elements of $A_\text{n-1}$ are equal to $1$ according to the assumption, and therefore it can be written as $A_\text{n-1}=\tilde{C}\tilde{C}^\dag$, where $\tilde{C}=(1,1,\cdots,1)^t$ is a $(n-1)$-dimensional column vector with all its components being $1$. Then, $A_\text{n-1}-C C^\dag $ can be written as $\tilde{C}\tilde{C}^\dag- C C^\dag$. Since $\tilde{C}\tilde{C}^\dag- C C^\dag$ is positive semidefinite, the average value of matrix operator $\tilde{C}\tilde{C}^\dag- C C^\dag$ with respect to any vector $\ket{\textbf{x}}=(x_1,x_2,\cdots,x_{n-1})^t$ must be nonnegative. It requires that $C=\tilde{C}=(1,1,\cdots,1)^t$, i.e., $a_{2,n}=\cdots=a_{n-1,n}=1$.
Otherwise, if some $a_{j,n}$ is not equal to $1$, we can always find a vector $\ket{\textbf{x}}$ with $x_1=1$, $x_j=-1$ and $x_{i\neq j}=0$, such that $\bra{\textbf{x}}(A_\text{n-1}-C C^\dag )\ket{\textbf{x}}<0$. This completes the proof of the above proposition.

\end{document}